\documentstyle[psfig]{mn}
\newcommand\etal{et al. }
\newcommand\refer{\par \noindent\hangindent=3pc \hangafter=1}

\newcommand\eg{eg }

\newcommand\ie{i.e. }

\title{An improved method of constructing binned luminosity functions}
\author[Page \& Carrera]{M.\,J. Page\(^{1}\), F.J. Carrera\(^{2}\)\\
\(^{1}\)Mullard Space Science Laboratory, University College London,
Holmbury St Mary, Dorking, Surrey RH5 6NT, UK.\\
\(^{2}\)Instituto de F\'\i sica de Cantabria (Consejo Superior de
Investigaciones Cient\'\i ficas--Universidad de Cantabria), 39005
Santander, Spain.}

\date{}

\begin{document}
\maketitle

\begin{abstract}

We show that binned differential
luminosity functions constructed using the \(1/V_{a}\) method
have a significant systematic error for objects close to their parent 
sample's flux limit(s).
This is particularly noticeable when luminosity functions are produced for
a number of different redshift ranges as is common in the study of AGN or
galaxy evolution.
We present a simple method of constructing a binned luminosity function 
which overcomes this problem and has a number of other advantages 
over the traditional  
\(1/V_{a}\) method. We also describe a practical method for comparing 
binned and model
luminosity functions, by calculating the expectation values of the binned 
luminosity function from the model. 

Binned luminosity functions produced by the two methods 
are compared for simulated data and for the Large Bright QSO Survey (LBQS). 
It is
shown that the \(1/V_{a}\) method produces a very misleading picture of
evolution in the LBQS.
The binned luminosity function of the LBQS is then compared to a model
two power law luminosity function undergoing pure luminosity evolution from
Boyle \etal (1991). The
comparison is made using a model
luminosity function averaged over each redshift shell, and using the
expectation values for the binned luminosity function calculated from the
model. The luminosity function averaged in each
redshift shell gives a misleading impression that the model over predicts the
number of QSOs at low luminosity even when model and data are consistent. 
The expectation values show that there are significant differences between 
model and data: the model 
overpredicts the
number of low luminosity sources at both low and high redshift. The luminosity
function does not appear to steepen relative to the model as
redshift increases.
 
\end{abstract}


\section{Introduction}

For three decades considerable effort has been spent in obtaining samples 
of AGN
to investigate their cosmological evolution. This evolution is
seen as a change in the luminosity function with redshift. 
In recent years,
galaxy redshift surveys and X--ray cluster surveys 
have become sufficiently deep that the evolution of galaxies and clusters of
galaxies
can be examined in the same way as that of AGN. 
Quantifying and understanding the evolution
of these populations is essential to modern cosmology.

A number of methods exist for demonstrating and quantifying the cosmological
evolution of a population, \eg the \(\langle V/V_{max}\rangle\) method of
Schmidt (1968) or by comparison with Monte Carlo simulations (La Franca \&
Cristiani 1997).
The simplest, most intuitive method is to construct a binned differential
luminosity function in a number of redshift intervals. For flux limited samples
(which are the norm in evolution studies), this is usually done using the
\(1/V_{a}\) method (\eg Maccacaro \etal 1991, Ellis \etal 1996). However, the
\(1/V_{a}\) method introduces a significant error for objects close to the flux
limit, and hence is {\it not} the optimum method for constructing binned
differential luminosity functions. We present a simple, alternative method
which we show is superior to \(1/V_{a}\).

This paper is laid out as follows. Section \ref{sec:method} 
describes both the \(1/V_{a}\)
method, and our new method,
for constructing binned differential luminosity functions. The relative
merits of the two methods
are compared in Section \ref{sec:advantages}, and in Section \ref{sec:sim}
a Monte Carlo simulation is
used to demonstrate the improvement that our method offers over the
\(1/V_{a}\) method. In Section \ref{sec:binmod} 
we present a technique for comparing binned
and model luminosity functions. 
We apply the two different
methods to construct binned luminosity functions of the Large Bright QSO
Survey sample in Section \ref{sec:lbqs}.
In Section \ref{sec:lbqsmod} this sample is compared to one of the model luminosity functions
and evolution laws from Boyle \etal (1991) using both an averaged luminosity
function and using the method given in Section \ref{sec:binmod}. 
Our conclusions are presented in Section \ref{sec:conclusions}.

\section{Method}
\label{sec:method}

\subsection{Fundamental quantities}

We define the differential luminosity function \(\phi\),
of any extragalactic population, as the number of objects 
per unit comoving volume per unit luminosity interval, \ie
\begin{equation}
\label{eq:phi1}
\phi(L,z)=\frac{d^{2}N}{dVdL}(L,z)
\end{equation}
where \(N\) is the number of objects of luminosity \(L\) 
found in comoving volume  \(V\) at redshift \(z\). 
We assume \(\phi\) is a continuous function over 
the range of redshift and luminosity for which it is defined.

The differential luminosity function is often defined as a function of  
logarithmic luminosity or (by optical astronomers) 
magnitude rather than luminosity. The method described here is 
equally applicable to luminosity functions defined in this way, 
with the appropriate substitution 
of `log \((L)\)' or `\(m\)' where we have used `\(L\)'.

\subsection{The \(1/V_{a}\) method}

The \(1/V_{a}\) method was originally proposed (Schmidt 1968)
to measure space 
density (\(dN/dV\)); 
the method was generalised for samples with multiple flux limits in
Avni \& Bahcall (1980).
If \(N\) objects have luminosities and redshifts in the interval 
\(\Delta L \Delta z\) around the bin centre \((L,z)\):
\[\frac{dN}{dV}(L,z) \approx \sum_{i=1}^{N} \frac{1}{V_{a}(i)}\]
where \(V_{a}(i)\) is the survey volume in which object \(i\) 
with luminosity \(L(i)\) could 
have been detected and remained in the redshift bin \(\Delta z\).
It is an unbiased estimator of the space density (Felten 1976); 
the main advantage of this method (compared with 
simply dividing the number of 
objects found by the average volume searched) is that it takes account 
of the fact that more luminous objects are detectable over a larger volume 
than the less luminous objects.

The \(1/V_{a}\) method is used to 
estimate \(\phi\) by dividing the space density 
by the luminosity  bin width. We will refer to the approximation of \(\phi\) 
obtained using this method as \(\phi_{1/V_{a}}\):
\begin{equation}
\label{eq:va}
\phi_{1/V_{a}}(L,z)=\frac{1}{\Delta L}\sum_{i=1}^{N} \frac{1}{V_{a}(i)}
\end{equation}

\subsection{A binned approximation to $\phi$}
\label{sec:approx}

The luminosity function as defined in equation \ref{eq:phi1} 
is related to the expected number 
of objects found in any region \(\Delta L \Delta V(\Delta z)\) 
of the volume-luminosity plane by 
\begin{equation}
\label{eq:expectation}
\langle N \rangle = \int_{L_{min}}^{L_{max}} 
\int^{z_{max}(L)}_{z_{min}} \phi(L,z) \frac{dV}{dz} dz dL
\end{equation}
where \(z_{min}\) is the bottom of the redshift interval \(\Delta z\) and
\(z_{max}(L)\) is the highest redshift possible for an object of 
luminosity L to be detected and remain in the redshift interval \(\Delta
z\). Angled  
brackets denote expectation value.
If \(\phi\) changes little over \(\Delta L \Delta z\), then
\begin{equation}
\label{eq:small}
\langle N \rangle \approx \phi(L,z) \int_{L_{min}}^{L_{max}}
\int^{z_{max}(L)}_{z_{min}}
\frac{dV}{dz} dz dL
\end{equation}
It follows that if \(N\) objects have been found over some  
volume-luminosity region \(A\),
\begin{equation}
\label{eq:approx}
\phi \approx \phi_{est} = \frac{N}{\int_{L_{min}}^{L_{max}}
\int^{z_{max}(L)}_{z_{min}}
\frac{dV}{dz} dz dL}
\end{equation}
where \(\phi_{est}\) is our binned estimate of the luminosity function.

\section{Comparison of the two methods}
\label{sec:advantages}

\subsection{Which is the better estimate of $\phi$ ?}
\label{sec:which}

\begin{figure*}
\begin{center}
\leavevmode
\psfig{figure=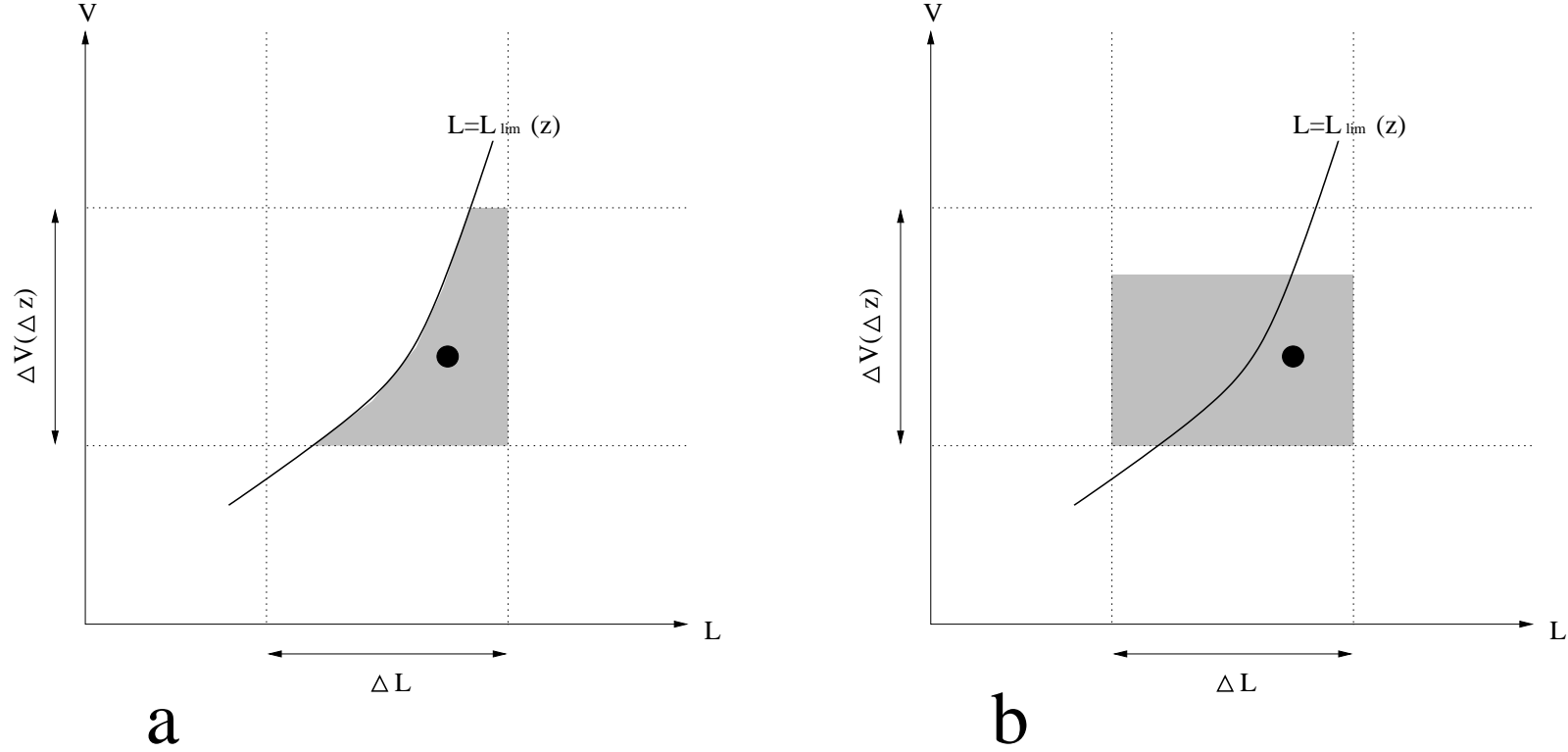,height=100truemm,width=180truemm,angle=270}
\caption{Volume-luminosity space `available' to an object (the black dot) in
(a) $\phi_{est}$ and (b) $\phi_{1/V_{a}}$. The line L=L$_{lim}$($z$) is the
minimum detectable luminosity of an object of redshift $z$ and is determined
by the flux limit of the survey.}
\label{fig:redlum}
\end{center}
\end{figure*}

Consider a single bin, corresponding to the  
redshift - luminosity
interval \(\Delta L\ \Delta z\) around the point (\(L_{1},z_{1}\)), in a binned
luminosity function constructed from a flux limited sample
of objects. 
We assume that
\(\Delta L\)  and \(\Delta z\) are sufficiently small and/or that \(\phi\)
is a sufficiently slowly varying function of \(L\) and \(z\) that our single
estimate of \(\phi\) is appropriate over the whole interval. 
Some portion of the region \(\Delta L\ 
\Delta z\) may represent objects which are fainter than the survey flux
limit. This situation is illustrated in Fig. \ref{fig:redlum} where the
curve \(L=L_{lim}(z)\) represents the flux limit of the survey.

The shaded region in Fig. \ref{fig:redlum}a is that region of the volume 
luminosity plane in the interval \(\Delta L\
\Delta z\) which has been surveyed. This shaded area 
corresponds to the double integral in equation \ref{eq:small}. 
By definition (equation \ref{eq:approx}), \(\phi_{est}\) 
gives a good estimate to
\(\phi(L_{1},z_{1})\). 

The shaded region in Fig. \ref{fig:redlum}b has an area equal to (\(\Delta L\
V_{a}(i)\)) for object i, represented by the black spot. This area is
clearly {\it not} the same as that of the surveyed region 
(the shaded region in Fig. \ref{fig:redlum}a) and hence 
there is no guarantee that \(\phi_{1/V_{a}}\)
will give a good estimate of \(\phi\). However, there are two specific cases
in which \(\phi_{1/V_{a}}\) will always give a good estimate of \(\phi\): 

\vspace{3mm}
\noindent
1) The entire redshift, luminosity interval \(\Delta L\ \Delta z\)
corresponds to objects brighter than the flux limit of the survey, \ie in 
Fig. \ref{fig:redlum} the curve \(L=L_{lim}(z)\) passes above the 
\(\Delta L\ \Delta z\) region. In this case,
\[\int \int dV dL=\Delta L\ \Delta V(\Delta z)\]
and
\[\phi_{1/V_{a}}=\phi_{est}\approx \phi\]

\vspace{3mm}
\noindent
2) \(\Delta L\) is very small: 
\[\phi_{1/V_{a}} \approx \frac{1}{\Delta L}
\frac{dN}{dV}\rightarrow \frac{d^{2}N}{dVdL}\ \ {\rm as}\ \ \Delta L 
\rightarrow 0\]

\vspace{3mm}
\noindent
Note that even in these limiting cases, \(\phi_{1/V_{a}}\) does not give a
better estimate to \(\phi\) than \(\phi_{est}\); instead \(\phi_{1/V_{a}}\)
and \(\phi_{est}\)  have the same value in case 1 and converge to the same
value in case 2. Case 1 generally applies for luminosity functions of 
objects which are much brighter than the flux limits. Hence for luminosity
functions in a number of redshift intervals, the two methods will produce
the same results for 
the highest luminosity bins of any given redshift shell. At the lowest
luminosities in each redshift shell, 
where objects are close to the survey flux limit and the shaded 
region
of volume-luminosity space in Fig.\ref{fig:redlum}a is much smaller than the
rectangular \(\Delta L \Delta V\) region, 
the two methods are likely to give the most different results.
In all cases, \(\phi_{est}\) can be expected to give as good or better
estimates for \(\phi\) than \(\phi_{1/V_{a}}\), and is therefore the better
estimator.

\subsection{Statistical uncertainty in the two methods}

From the definition of \(\phi_{est}\) in Section \ref{sec:approx} it is
easily shown that the statistical uncertainty, \(\delta \phi_{est}\), is
given by:
\begin{equation}
\delta \phi_{est}=\frac{\delta N}{\int_{L_{min}}^{L_{max}}
\int^{z_{max}(L)}_{z_{min}}\frac{dV}{dz} dz dL}
\label{eq:approxerror}
\end{equation}
where \(\delta N\) is the uncertainty on \(N\) objects and can be calculated
from Poisson or Gaussian statistics as appropriate.

The statistical uncertainty on \(\phi_{1/V_{a}}\) is harder to estimate,
because each object makes a different statistical contribution. 
The uncertainty is typically estimated by the following expression: 
\begin{equation}
\delta \phi_{1/V_{a}}=\frac{1}{\Delta L} \sqrt{\sum_{i=1}^{N}
\Bigl( \frac{1}{V_{a}(i)}\Bigr)^2}
\end{equation}
(\eg Marshall 1985, Boyle \etal 1988) but this assumes Gaussian
statistics and therefore is not appropriate for bins containing few objects.

Hence the uncertainty on \(\phi_{est}\) is easily calculated for any number
of objects within a bin, but the uncertainty on \(\phi_{1/V_{a}}\) can only be
properly estimated when there are many sources per bin.

\subsection{Multiple flux limits}
\label{sec:multiple}

The comparison so far and Fig. \ref{fig:redlum} has assumed the simple case of
a survey with a single flux limit. However, many surveys have different flux
limits for different parts of the survey area, and most luminosity function
investigations combine different surveys with different flux limits. Indeed,
combining large area shallow surveys and small area deep surveys is
currently the only practical way to obtain a luminosity function which spans
a wide range of luminosities at a wide range of redshifts.
The most efficient means to combine areas with different flux limits is to
assume that each object could be found in any of the survey areas for which it
is brighter than the corresponding flux limit. This is 'coherent' addition of
samples in the language of Avni \& Bahcall (1980). In this case, it is no
longer true that only the lowest luminosity objects in each redshift shell are
close to the flux limit, because objects of higher luminosity are close to the
flux limit of the shallower survey areas. The problem with \(\phi_{1/V_{a}}\)
therefore affects luminosity bins that correspond to any of the survey 
flux limits, although the severity of the problem is watered down by the 
deeper survey areas.

Many surveys also have a {\it maximum} flux limit beyond which sources are too
bright to have been selected.  In this case \(z_{min}\) in equations
\ref{eq:phi1} - \ref{eq:approxerror}, as well as \(z_{max}\), is a function of
L. As an illustration, if the curve \(L=L_{lim}(z)\) in Fig. \ref{fig:redlum}
were to represent a maximum (rather than a minimum) flux limit, objects would
have to lie on the other side of the curve to be included in the survey. If the
object (black dot) in Fig. \ref{fig:redlum} were shifted straight upwards to
lie above the black line, it is now the shaded area of Fig. \ref{fig:redlum}a
subtracted from \(\Delta L\ \Delta V (\Delta z)\) which corresponds to the
double integral in equation \ref{eq:small}, and the shaded area of
Fig. \ref{fig:redlum}b subtracted from \(\Delta L\ \Delta V (\Delta z)\).
which corresponds to (\(\Delta L\ V_{a}(i)\)). Just as for the minimum flux
limit (section \ref{sec:which}), these two quantities are different, and there
is no guarantee that \(\phi_{1/V_{a}}\) will give a good estimate of
\(\phi\). Therefore, for surveys with both bright and faint flux limits,
\(\phi_{1/V_{a}}\) is likely to be inferior to \(\phi_{est}\) at the highest as
well as the lowest luminosity bins in each redshift shell.

\section{A simulated luminosity function}
\label{sec:sim}

\begin{figure*}
\begin{center}
\leavevmode
\psfig{figure=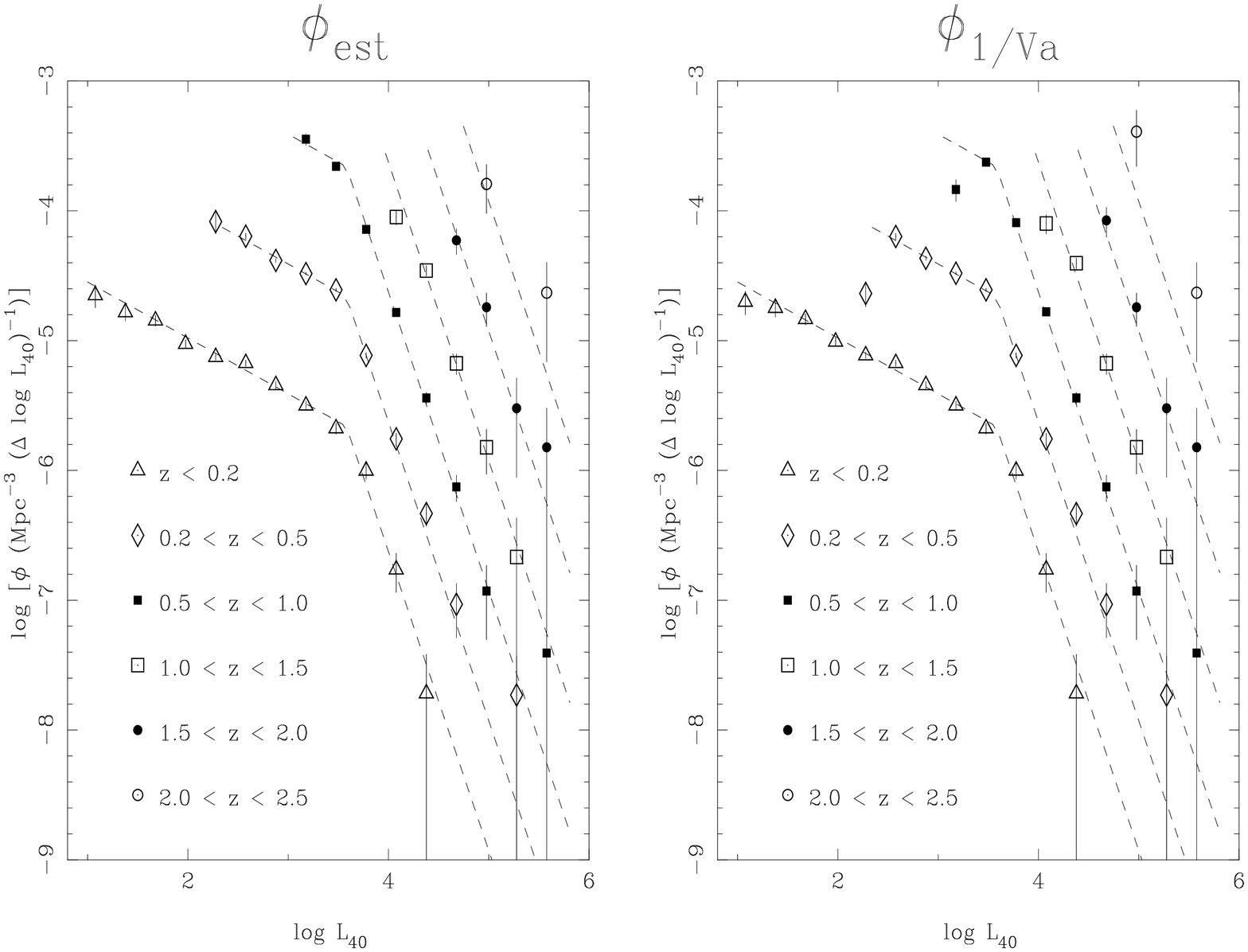,height=120truemm,width=180truemm,angle=0}
\caption{Binned luminosity functions of a simulated sample of objects using
(left) $\phi_{est}$ and (right) $\phi_{1/V_{a}}$. The luminosity functions
of successive redshift shells have been shifted in the vertical direction
for clarity; the input model has no evolution with redshift and is shown as
dashed lines.}
\label{fig:bothsim}
\end{center}
\end{figure*}

In this section we use a sample simulated by a Monte Carlo method to 
demonstrate the improvement that \(\phi_{est}\) offers over the
traditional \(\phi_{1/V_{a}}\). The simulation was performed using a two
power law model X--ray luminosity function which is unchanging with redshift
(\ie no evolution)
and a single flux limit.
The Monte Carlo simulation produced a source list of \(\sim 10000\) sources.
Binned luminosity functions 
were produced for this simulated data in a
range of redshift intervals using both methods.
These are shown in Fig. \ref{fig:bothsim}, \(\phi_{est}\) on the left and
\(\phi_{1/V_{a}}\) on the right. The model luminosity function is shown as a
dashed line. 

The two methods are in exact agreement for the 
high luminosity points of each redshift interval, as expected from Sec
\ref{sec:which}. However, for the lowest luminosity bins of the \(0.2<z<0.5\)
and \(0.5<z<1.0\) redshift intervals \(\phi_{1/V_{a}}\) is significantly
smaller than the input model, 
while \(\phi_{est}\) is a good representation of the input model. For the
lowest luminosity bin of the \(2.0<z<2.5\) redshift shell \(\phi_{1/V_{a}}\)
is larger than the input model.
We therefore see that for objects close to the flux limit 
\(\phi_{1/V_{a}}\) can be too small or it can be too large; in simulations
we have performed using different flux limits and luminosity functions 
\(\phi_{1/V_{a}}\) is more frequently too small than too large.

\section{Comparing binned and model luminosity functions}
\label{sec:binmod}

The luminosity function is frequently modelled as some analytical function (\eg
a broken power law or a Schechter function). Models are sometimes fitted to the
binned luminosity function and tested for goodness of fit using
\(\chi^{2}\). More often the fit is made to the unbinned data by maximum
likelihood (Crawford, Jauncey \& Murdoch 1970) and models are tested for
overall goodness of fit using a one or two dimensional Kolmogorov Smirnov test
(Press \etal 1992). However, even in this latter case, comparison between the
binned and model luminosity function is the usual recourse to find out {\it
why} models are rejected, \ie at which redshifts and luminosities the data and
model are inconsistent.

In this comparison, a problem can arise that the model is a continuous function
and (for a steep luminosity function) can span a rather large range of
values within any one luminosity bin. 
Similarly, any model luminosity
function which is evolving with redshift can have a large range of values
for a single luminosity over a redshift interval (although this latter effect
can be overcome if the evolution is known or assumed a-priori, \eg 
Mathez \etal 1996, Kassiola \& Mathez 1990).
The binned luminosity
function could be compared to the model luminosity function evaluated at
some arbitrary `middle point' of each luminosity/redshift bin; alternatively
some weighted average of the model could be produced for each luminosity
redshift bin.

A more valid statistical approach is to compute the expectation value of the
binned luminosity function from the model, for each
redshift/luminosity bin. In this method, a single, unique, model value is
produced for each redshift/luminosity bin, which can be compared to the
corresponding binned luminosity function data point. This allows the
binned luminosity function to be compared with the model using a statistical
goodness of fit test such as \(\chi^{2}\), and/or enables the fitting of a
model to the binned luminosity function. This approach is illustrated in
this section  using
the Monte Carlo simulated sample of section \ref{sec:sim}. First, the
formulas for expectation value are given.

\begin{boldmath}
\subsection{$\langle \phi_{1/V_{a}} \rangle$}
\end{boldmath}

The expectation value of $\phi_{1/V_{a}}$ is given by:
\begin{equation}
\langle \phi_{1/V_{a}}\rangle =\frac{\langle N \rangle}{\Delta L}
\int^{L_{max}}_{L_{min}}\frac{1}{V_{a}(L)}P(L)dL
\end{equation}
where \(L_{min}\) and \(L_{max}\) are respectively the minimum and maximum 
luminosities of objects within the luminosity bin and \(P(L)\) is the
probability density corresponding to an object of luminosity \(L\).
\(P(L)\) is given by

\begin{equation}
P(L)=\frac{1}{\langle N \rangle}
\int^{z_{max}(L)}_{z_{min}}\phi(L,z)\frac{dV}{dz} dz
\end{equation}

where \(z_{min}\) is the lower limit of the redshift shell and \(z_{max}(L)\)
is the smaller of the maximum detectable redshift of an object of luminosity
L and the top of the redshift shell. \(\langle N \rangle\) is given by equation
\ref{eq:expectation}. Hence 

\begin{equation}
\label{eq:modpredva}
\langle \phi_{1/V_{a}}\rangle =\frac{1}{\Delta L}
\int^{L_{max}}_{L_{min}}\frac{1}{V_{a}(L)}
\int^{z_{max}(L)}_{z_{min}}\phi(L,z)\frac{dV}{dz} dz\ dL
\end{equation}

\begin{boldmath}
\subsection{$\langle \phi_{est} \rangle$}
\end{boldmath}

From equations \ref{eq:expectation} and \ref {eq:approx}, 
the expectation value of
\(\phi_{est}\) is given by: 
\begin{equation}
\label{eq:modpredest}
\langle \phi_{est}\rangle =\frac{\int_{L_{min}}^{L_{max}}
\int^{z_{max}(L)}_{z_{min}} \phi \frac{dV}{dz} dz dL}{\int_{L_{min}}^{L_{max}}
\int^{z_{max}(L)}_{z_{min}}\frac{dV}{dz} dz dL}
\end{equation}

\begin{boldmath}
\subsection{$\langle \phi_{1/V_{a}} \rangle$ and $\langle \phi_{est}
\rangle$ for the simulated luminosity function}
\end{boldmath}
\label{sec:expsim}

\begin{figure*}
\begin{center}
\leavevmode
\psfig{figure=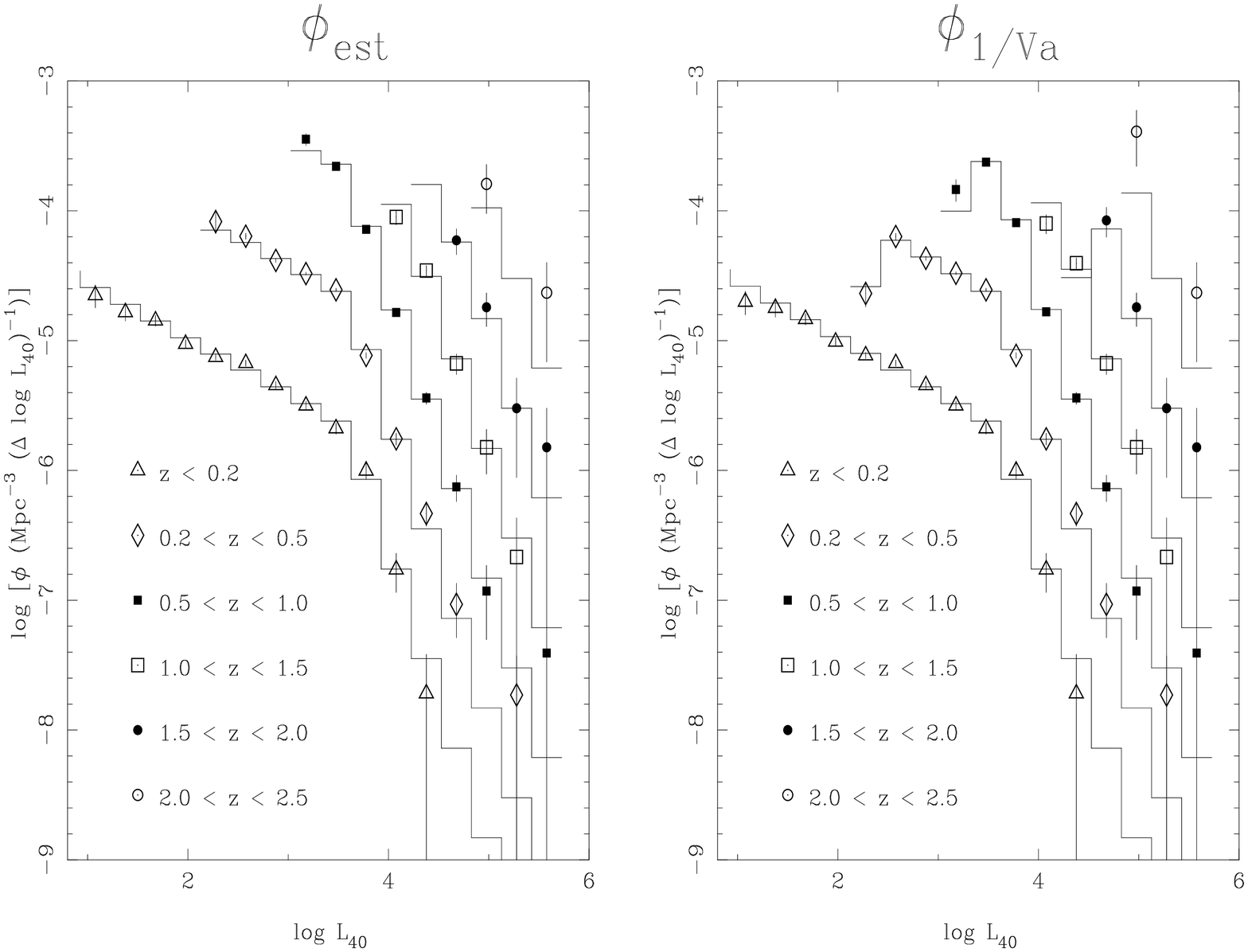,height=120truemm,width=180truemm,angle=0}
\caption{(left): $\phi_{est}$ (data points) and $\langle \phi_{est}\rangle$
(stepped line) and (right): $\phi_{1/V_{a}}$ (data points) $\langle
\phi_{1/V_{a}}\rangle$ (stepped line) for the Monte Carlo simulated sample.}
\label{fig:bothexp}
\end{center}
\end{figure*}

In Fig. \ref{fig:bothexp} \(\langle \phi_{1/V_{a}}\rangle\) and 
\(\langle \phi_{est}\rangle\) are shown with \(\phi_{1/V_{a}}\) and
\(\phi_{est}\) of the Monte Carlo simulated sample of section \ref{sec:sim}.
Unsurprisingly, the binned luminosity function of the Monte Carlo simulated
sample is consistent with the expectation luminosity function derived from
the input model.
Note that the 
systematic differences between \(\phi_{1/V_{a}}\) and \(\phi\) 
(the turn down of \(\phi_{1/V_{a}}\) at low luminosities in Fig.
\ref{fig:bothsim}) are reproduced by  
\(\langle \phi_{1/V_{a}}\rangle\). Hence it is in principle possible to
compare and/or fit a model \(\phi\) to a binned \(\phi_{1/V_{a}}\) provided
that the model is first transformed into \(\langle \phi_{1/V_{a}}\rangle\).
However, it is simpler to evaluate equation \ref{eq:modpredest} for arbitrary 
\(\phi(L,z)\)
than equation \ref{eq:modpredva}, so we are again led to the conclusion that 
\(\phi_{est}\) is superior to \(\phi_{1/V_{a}}\).

\section{The luminosity function of the Large Bright QSO survey}
\label{sec:lbqs}

\begin{figure*}
\begin{center}
\leavevmode
\psfig{figure=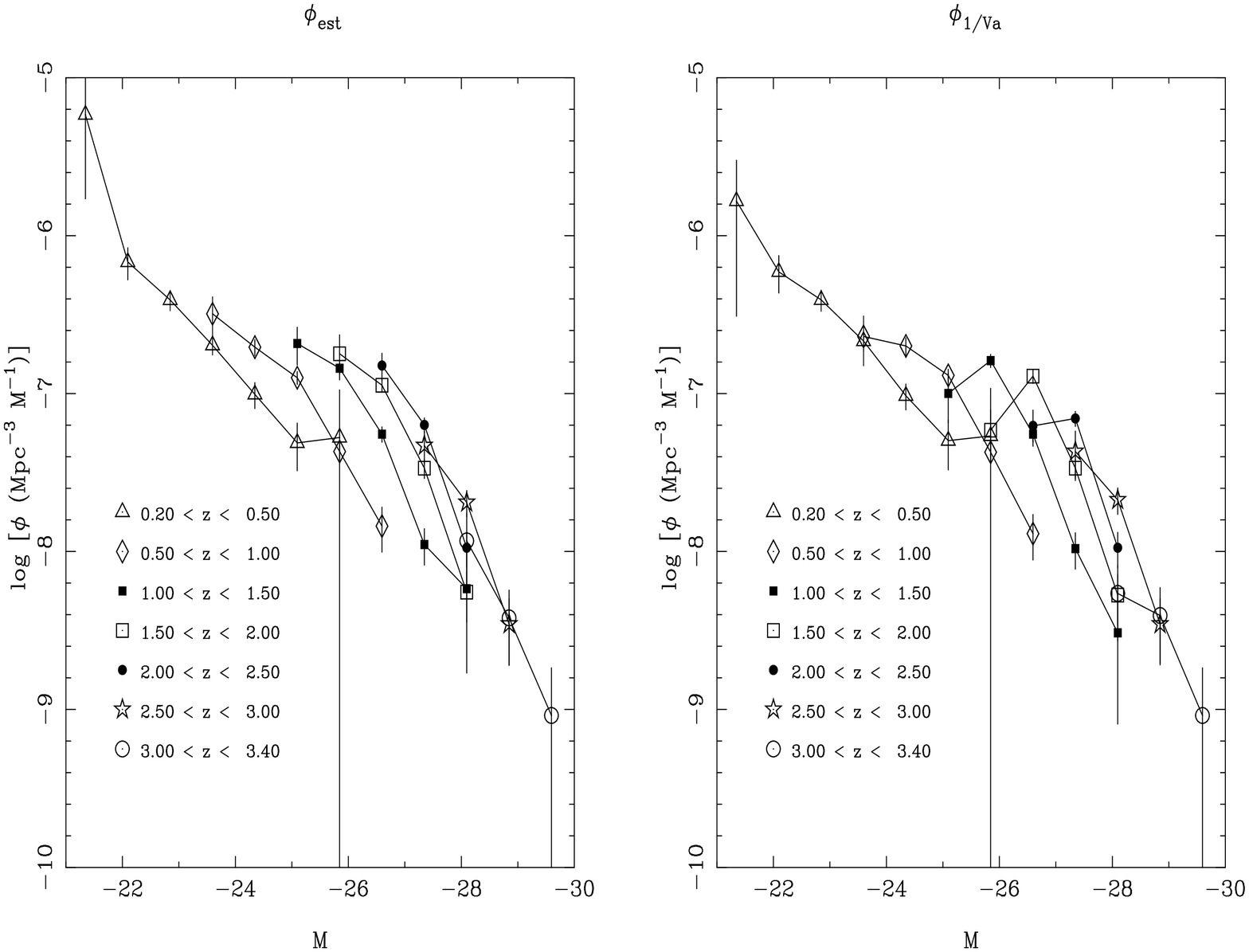,height=120truemm,width=180truemm,angle=0}
\caption{Binned luminosity functions of the LBQS sample: (left) $\phi_{est}$
and (right) $\phi_{1/V_{a}}$}
\label{fig:lbqs}
\end{center}
\end{figure*}

To demonstrate the difference between binned luminosity functions produced
using \(\phi_{est}\) and \(\phi_{1/V_{a}}\), we now apply these two methods to
real data.  For this we have chosen the Large Bright QSO Survey (hereafter
LBQS, Hewett, Foltz \& Chaffee 1995), the current largest single sample of QSOs
available.  This is ideal because as samples become larger, the systematic
problems with \(\phi_{1/V_{a}}\) become larger than the statistical uncertainty
associated with the data.  Note that the LBQS has a maximum flux limit as well
as minimum flux limits (see section \ref{sec:multiple}); these are given by
Hewett \etal (1993).

\(\phi_{est}\) and \(\phi_{1/V_{a}}\) are shown for the LBQS in Fig.
\ref{fig:lbqs} for \(q_{0}=0.5\) and H\(_{0}\)=50 km s\(^{-1}\) Mpc\(^{-1}\).
K-correction was performed using the composite spectrum of Cristiani \& Vio
(1990).  The contrast is striking: the \(\phi_{1/V_{a}}\) binned luminosity
function gives the misleading impression that evolution is complex and highly
luminosity dependent while \(\phi_{est}\) shows a much simpler picture of the
evolution; the latter could be described approximately by pure luminosity
evolution.

\section{Comparison of the Large Bright QSO Survey 
with a model luminosity function}
\label{sec:lbqsmod}

\begin{figure*}
\begin{center}
\leavevmode

\psfig{figure=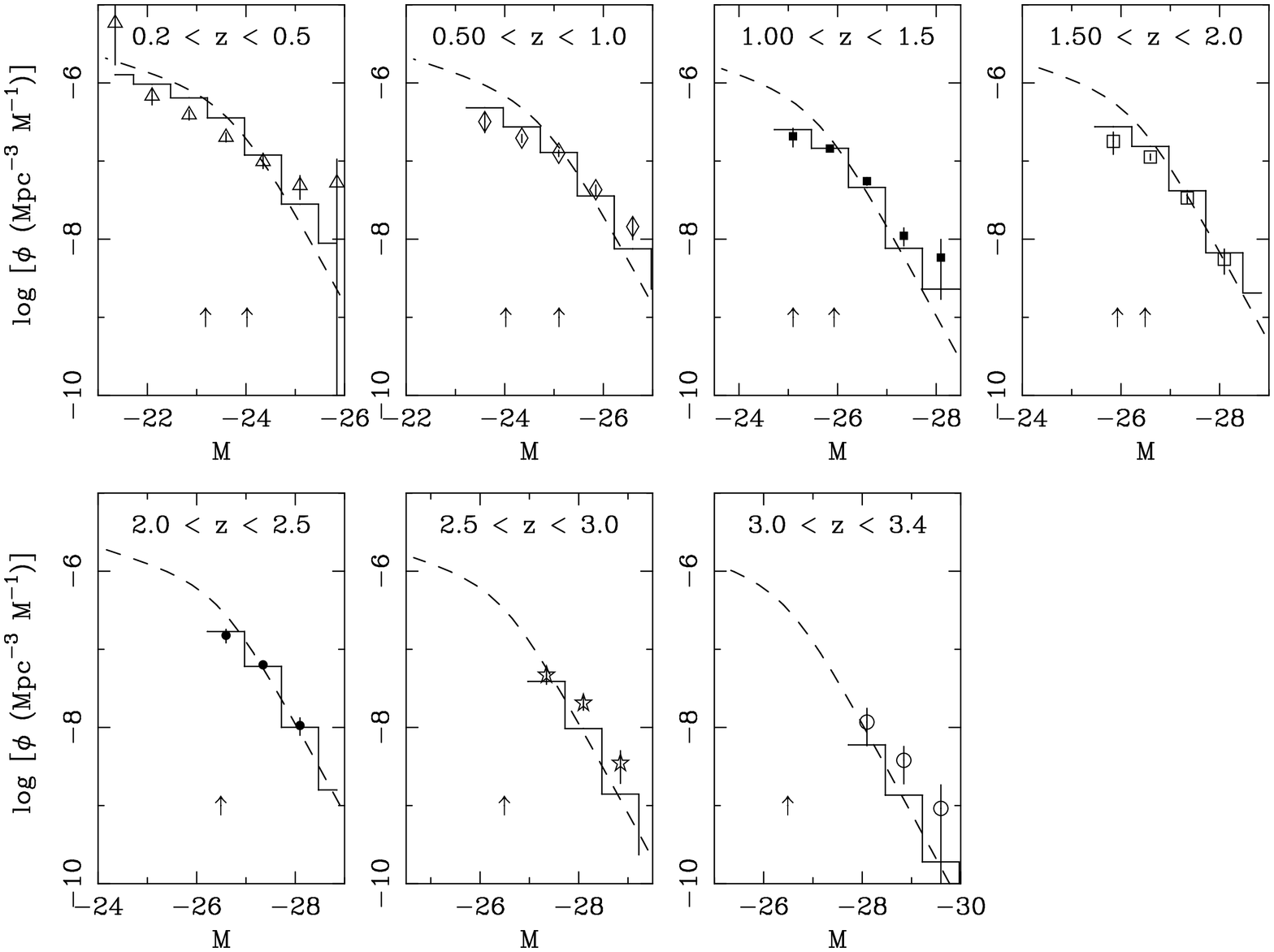,height=120truemm,width=180truemm,angle=270}
\caption{$\phi_{est}$ for the LBQS (data points) in different redshift 
shells, with an average model luminosity function 
(continuous dashed line, see Sec: \ref{sec:lbqsmod})
and $\langle \phi_{est}\rangle$ for the LBQS and the same model
(stepped solid line).}
\label{fig:lbqs7wins2}
\end{center}
\end{figure*}

We now compare the binned luminosity function \(\phi_{est}\) of the LBQS 
with a model
luminosity function undergoing pure luminosity evolution. The model chosen is 
the \(q_{0}=0.5\) model from Boyle \etal (1991), which is a two power law
luminosity function described by: 
\[\phi(M_{B_{J}})=\phi^{*} 
\{ 
10^{0.4 [M_{B_{J}}-M_{break}(z)](\alpha + 1)} \]
\[\ \ \ \ \ \ +10^{0.4 [M_{B_{J}}-M_{break}(z)](\beta + 1)} 
\}^{-1}\]
where \(\phi^{*}=6.5\times 10^{-7}\) mag\(^{-1}\) Mpc\(^{-3}\),
\(\alpha=-3.9\) and \(\beta=-1.5\). The model is subject to pure luminosity
evolution such that \(M_{break}(z)\)=-22.5 - 8.625 log (1+\(z\)), until
\(z=1.9\) after which \(M_{break}(z)\) remains constant at \(M_{break}(1.9)\).
The binned luminosity function is shown in Fig. \ref{fig:lbqs7wins2};
the different redshift shells have been placed in separate panels for clarity.
For comparison the model is shown as both a 
volume-weighted average over each redshift shell (dashed line), 
and as a set of expectation
values 
\(\langle \phi_{est}\rangle\) calculated using Equation \ref{eq:modpredest}
(solid stepped line). The two arrows near the bottom of each plot show the
position of the `break' luminosity of the model luminosity function 
at the top and bottom of the redshift shell. 

The advantage of comparing the binned luminosity function to expectation values
is well 
demonstrated by
the low luminosity bins of the upper four panels of Fig. \ref{fig:lbqs7wins2},
in which the continuous volume weighted luminosity function 
makes the data look far more discrepant than it really is. 

The model from Boyle \etal (1991) has already been compared to luminosity
function of the LBQS, by Hewett \etal (1993).  These authors used cumulative
1/Va luminosity functions in the same redshift shells as used here; these were
compared to the model predictions using Kolmogorov Smirnov tests in each
redshift shell. Their most important conclusion was that at \(z<2\) the
luminosity function systematically steepens relative to the model with
increasing \(z\), implying that pure luminosity evolution is no longer a viable
model for QSO evolution. Similar claims for other surveys have been made more 
recently by 
Goldschmidt \& Miller (1998), La Franca \& Cristiani (1997) and by Hawkins \& Veron (1995).

The binned luminosity function of Fig. \ref{fig:lbqs7wins2} does not support
this conclusion. Instead, significant discrepancy between data and model at
\(z<2\) is always for fainter absolute magnitudes, at or below
\(M_{break}(z)\), and is always in the sense that the model overpredicts the
number of low luminosity sources. To quantify this, we calculated the
\(\chi^{2}\) separately for bins brighter, and for bins fainter
\(M_{break}(z)\) in the four redshift shells at \(z<2\). These are tabulated in
Table \ref{tab:chisq}; the lowest and highest luminosity bins of the
\(0.2<z<0.5\) redshift shell and the highest luminosity bin of the
\(1.0<z<1.5\) redshift shell have not been included because they contain too
few objects for \(\chi^{2}\) to be appropriate.  At absolute magnitudes
brighter than \(M_{break}(z)\) the model is acceptable at 95\%, overall and for
each redshift shell individually. At fainter absolute magnitudes the model is
significantly deviant for all but the \(1.0<z<1.5\) redshift shell, and it is
notable that in this \(1.0<z<1.5\) shell the model shows the largest deficiency
relative to the data at high luminosities. Although the discrepancy between
data and model at low luminosities is sufficient to reject the model at
\(>7\sigma\), there is no strong evidence that the
luminosity function steepens relative to the model as redshift increases. 

\begin{table}
\caption{Model goodness of fit at absolute magnitudes brighter, and similar to
or fainter than,
$M_{break}(z)$ for $z<2$}
\label{tab:chisq}
\begin{tabular}{ccccc} 
&\multicolumn{2}{c}{$M_{B_{J}}<M_{break}(z)$}
&\multicolumn{2}{c}{$M_{B_{J}} \geq M_{break}(z)$}\\
Redshift shell&$\chi^{2}/\nu$&P$^{*}$&$\chi^{2}/\nu$&P$^{*}$\\
&&&&\\
$0.2<z<0.5$&2.89/2&0.24&58.70/3&1.1$\times 10^{-12}$\\
$0.5<z<1.0$&3.61/2&0.16&14.72/3&2.1$\times 10^{-3}$\\
$1.0<z<1.5$&3.72/2&0.16&0.68/2&0.71\\
$1.5<z<2.0$&3.30/2&0.19&19.21/2&6.7$\times 10^{-5}$\\
Total     &13.52/8&0.10&93.31/10&1.2$\times 10^{-15}$\\
&&&&\\
\multicolumn{5}{l}{* Probability of $\chi^{2}$ being higher by chance}\\
\multicolumn{5}{l}{\ \ \ if the data were drawn from the model}\\
\end{tabular}
\end{table}

\section{Conclusions}
\label{sec:conclusions}

We demonstrate that the \(1/V_{a}\) method can lead to systematic errors when
used to produce binned differential luminosity functions. This problem is most
significant for objects close to their parent sample's flux limit(s). As a
result the lowest luminosity bins of \(1/V_{a}\) luminosity functions which are
split by redshift can be unrepresentative, distorting the apparent evolution of
extragalactic populations.  A new method for constructing binned luminosity
functions, which does not have this problem, is presented. The improvement of
this new method over the \(1/V_{a}\) method is demonstrated using a Monte Carlo
simulated sample of objects. This new method also has the advantages that 
statistical uncertainty is easily estimated even when there are few objects per
bin. 

We also present a practical method for comparing binned and model luminosity
functions (by eye or by statistical test) 
which resolves the problem that the model luminosity function has
many values within one luminosity-redshift bin.

We demonstrate the difference between the \(1/V_{a}\) method and 
the new method for constructing binned luminosity functions with
the Large Bright QSO Survey sample. Evolution appears complex and highly
luminosity dependent when the \(1/V_{a}\) method is used, but
relatively simple in the binned luminosity function produced 
using the new method.
We also use the LBQS, along with a model luminosity function and evolution law
from Boyle \etal (1991), to demonstrate the advantages of our method for 
comparing model and binned luminosity functions. We show that the model
is inconsistent with the data at low luminosities, but unlike 
Hewett \etal (1993) we do not find strong evidence that the
luminosity function steepens relative to the model as
redshift increases.

\section{Acknowledgments}

We would like to thank P. Hewitt for supplying the LBQS source list in
electronic form and for very useful comments that have improved this paper. 
FJC thanks the DGES for partial financial support, under project PB95-0122.

\section{References}

\refer Avni Y., and Bahcall J.N., 1980, ApJ, 235, 694

\refer Boyle B.J., Shanks T., Peterson B.A., 1988, MNRAS, 235, 935

\refer Boyle B.J., Jones L.R., Shanks T., Marano B., Zitelli V., Zamorani G., 
1991, The Space Distribution of Quasars, ASP Conference Series 21, 191

\refer Cristiani S., \& Vio R., 1990, A\&A, 227, 385

\refer Crawford D.F., Jauncey D.L. \& Murdoch H.S., 1970, ApJ, 162, 405

\refer Ellis R.S., Colless M., Broadhurst T., Heyl J., Glazebrook K., 1996,
MNRAS, 280, 235

\refer Felten J.E., 1976, ApJ, 207, 700

\refer La Franca F., Cristiani S., 1997, AJ, 113, 1517 

\refer Goldschmidt P., \& Miller L., 1998, MNRAS, 293, 107

\refer Hawkins M.R.S., \& Veron P., 1995, MNRAS, 275, 1102

\refer Hewett P.C., Foltz C.B., Chaffee F.H., 1993, ApJ, 406, L43

\refer Hewett P.C., Foltz C.B., \& Chaffee F.H., 1995, AJ, 109, 1498

\refer Kassiola A. \& Mathez G., 1990, A\&A, 230, 255

\refer Maccacaro T., Della Ceca R., Gioia I.M., Morris S.L., Stocke J.T.,
Wolter A., 1991, ApJ, 374, 117

\refer Marshall H.L., 1985, ApJ, 299, 109

\refer Mathez G., Van Waerbeke L., Mellier Y., Bonnet H., Lachie\`ze-Rey M.,
1996, A\&A, 316, 19 

\refer Press W.H., Teukolsky S.A., Vetterling W.T., Flannery B.P., 1992,
Numerical Recipes in Fortran, Cambridge University Press

\refer Schmidt M., 1968, ApJ, 151, 393

\end{document}